% PLAIN TEX
\magnification1200
\def\sk{Sherrington-Kirkpatrick}
\def\fl{fluctuations}
{\settabs 2\columns
\+Preprint n. &Dipartimento di Fisica\cr
\+&Universit\`a di Roma ``La Sapienza''\cr
\+&I.N.F.N. - Sezione di Roma\cr}
\vskip 2cm
\centerline{\bf FLUCTUATIONS AND THERMODYNAMIC VARIABLES}
\centerline{{\bf IN MEAN FIELD SPIN GLASS MODELS}\footnote{$\ddag$}{Research
supported in part by MURST (Italian Minister of University and Scientific and
Technological Research) and INFN (Italian National Institute for Nuclear
Physics).}} 
\smallskip 
\centerline{by}
\smallskip \centerline
{Francesco Guerra}
\centerline{Dipartimento di Fisica, Universit\`a di Roma ``La Sapienza'',}
\centerline{and Istituto Nazionale di Fisica Nucleare, Sezione di Roma,}
\centerline{Piazzale Aldo Moro, 2, I-00185 Roma, Italy.}
\centerline{e-mail guerra@roma1.infn.it}
\bigskip\bigskip
\centerline{April 1992}
\vfill
\beginsection ABSTRACT.

We present two rigorous results on the Sherrington-Kirkpatrick mean field model
for spin glasses, proven by elementary methods, based on properties of
fluctuations, with respect to the external quenched noise, of the thermodynamic
variables and order parameters. The first result gives the {\sl uniform}
convergence of the quenched average of the free energy in the thermodynamic
limit to its annealed approximation, in the high temperature regime, {\sl
including} the assumed critical point ($\beta=1$ in our notations). The second
result shows that the free energy can be expressed through a functional order
parameter, of the type introduced by Parisi in the frame of the replica
symmetry breaking method. The functional order parameter is implicitely given
in terms of fluctuations of thermodynamic variables.
\vfill\eject

\beginsection 1. INTRODUCTION.

The Sherrington-Kirkpatrick mean field model [1] for spin glasses is defined
through the Hamiltonian
$$H_N(\sigma,J)=-{1\over\sqrt{N-1}}\sum_{(i,j)}J_{ij}\sigma_i\sigma_j,
\eqno(1.1)$$
where $\sigma_1,\dots,\sigma_N$ are Ising spins, with values $\pm1$, coupled in
a highly disordered way through the quenched variables $J_{ij}$, assumed to be
independent random variables with identical unit Gaussian distribution (for the
sake of simplicity). The sum runs over all the $N(N-1)/2$ couples $(i,j)$ of
spins.

Due to its relevant physical interest as a prototype for disordered complex
systems, this model has been the subject of intensive investigation in the last
years [2]. By now, its general structure is well understood, at least from a
qualitative point of view, and ingeniously described by the Parisi solution
[3], originally obtained through the replica symmetry breaking method, and
further confirmed by the cavity method [4]. According to the accepted view, the
model is trivial in the high temperature regime ($\beta\le1$), where the
thermodynamic variables coincide with their annealed approximation, in the
thermodynamic limit $N\to\infty$. On the other hand, at low temperatures
($\beta>1$), the model exhibits an extremely rich structure of approximate
equilibrium states, separated by high barriers, and organized in a
hierarchical system. As a consequence, the thermodynamic variables are
expressed through a functional order parameter, introduced by Parisi [5],
replacing the numerical order parameter of the approximate 
Sherrington-Kirkpatrick solution [1], obtained without replica symmetry
breaking and violating positivity of the entropy.

Rigorous results, by Aizenman, Lebowitz and Ruelle [6], give a very detailed
description of the fluctuations of thermodynamic variables and their limiting
behavior, as $N\to\infty$, in the high temperature regime ($0\le\beta<1$)
(see also [7]), and interesting estimates on the free energy at low
temperature.

On the other hand, Pastur and Scherbina [8] have given a rigorous proof of
the damping of the mean square fluctuations of the free energy, in the
thermodynamic limit, at {\sl all} fixed temperatures. They have also shown
that the approximate \sk\ solution, surely defective at high $\beta$, follows
necessarily from the assumption that a suitably defined order parameter,
connected with the induced magnetization in an external random field, has
vanishing mean square \fl\ in the thermodynamic limit (see also [9]). This
result strongly supports Parisi theory.

The study of \fl\ is clearly of great importance for this model. In this
paper, we give a rigorous proof of two results, based on elementary
properties of \fl\ with respect to the external quenched noise $J$. The first
result gives the {\sl uniform} convergence of the quenched average of the
free energy to its annealed approximation, in the high temperature regime, {\sl
including} the  point $\beta=1$, which is the critical point, according to
the accepted picture. The second result is based on the evaluation of the
corrections necessary to go from the quenched average to the annealed
average, where the variables $J$ participate to the thermodynamic
equilibrium. These corrections are expressed through \fl\ of appropriate     
thermodynamic variables. A simple consequence is the emergency of the
functional order parameter and the antiparabolic martingale equation of Parisi
theory [3]. Therefore, we show that the free energy can be expressed through
the functional order parameter, exactly as in Parisi theory. For further
developments along this line we refer to [10].

The paper is organized as follows. In Section 2, we give a detailed description
of the model and some elementary properties of the thermodynamic variables.
The results on the high temperature behavior are given in Section 3, where we
exploit only simple convexity properties and positivity of \fl. In Section 4
we introduce the marginal martingale method, equivalent to the cavity method.
It is a very powerful technique for the study of the thermodynamic limit in
mean field models.

The functional order parameter and the antiparabolic martingale equation are
introduced in Section 5. They allow a very simple expression of the
thermodynamic variables of the system. The functional order parameter can be
easily written in terms of \fl, which take into account the corrections
necessary to go from  quenched averages to  annealed averages.

Finally, Section 6 is devoted to conclusions and outlook for future
developments.

\beginsection 2. THE MODEL. ELEMENTARY PROPERTIES.

We consider $N$ sites, $i=1,\dots,N$. In the thermodynamic limit we will let
$N\to\infty$. The configurations of the system are given by
$$\sigma:\quad\{1,2,\dots,N\}\ni i\to\sigma_i\in Z_2=\{-1,1\},
\eqno(2.1)$$
where $\sigma_i$ are Ising spins. For each of the $N(N-1)$ couples of sites
$(i,j)$, $i\ne j$, we introduce independent random variables $J_{ij}=J_{ji}$, $i\ne
j$,  identically distributed , called quenched variables. The
$\sigma$'s are mesoscopic random variables subject to thermodynamic
equilibrium. The $J$'s do not participate to thermodynamic equilibrium,
but act as a kind of random environment on the $\sigma$'s. For the sake
of simplicity, we assume that the $J$'s have unit Gaussian distribution
with
$$E(J_{ij})=0,\quad E(J_{ij}^2)=1,
\eqno(2.2)$$
where $E$ denotes averages with respect to the $J$ variables.

The Hamiltonian of the model is given by
$$H_N(\sigma,J)=-{1\over\sqrt{N-1}}\sum_{(i,j)}J_{ij}\sigma_i\sigma_j,
\eqno(2.3)$$
where the sum extends over all the $N(N-1)$ distinct couples. The square
root is necessary in order to assure the right thermodynamic behavior
for the extensive variables, as it will be clear in the following. We
write it in the form $\sqrt{N-1}$, equivalent to the customary one given
by $\sqrt{N}$. Introducing the inverse temperature $\beta$ (in proper
units), we can write the {\sl Boltzmannfaktor} as $\exp(-\beta
H_N(\sigma,J))$, and define the partition function $Z_N(\beta,J)$ and the
free energy $F_N(\beta,J)$ through
$$Z_N(\beta,J)=\sum_{\sigma_1\dots\sigma_N}\exp(-\beta
H_N(\sigma,J))=\exp(-\beta F_N(\beta,J)).
\eqno(2.4)$$
The associated Boltzmann state $\omega_{N,\beta,J}$ is given by
$$\omega_{N,\beta,J}(A)=Z_N(\beta,J)^{-1}\sum_{\sigma_1\dots\sigma_N}A\exp(-\beta
H_N(\sigma,J)), 
\eqno(2.5)$$
where $A$ is a generic function of the $\sigma$'s. Its density is
$$\rho_N(\sigma;\beta,J)=Z_N(\beta,J)^{-1}\exp(-\beta H_N(\sigma,J)). 
\eqno(2.6)$$
The internal energy $U_N(\beta,J)$ is
$$U_N(\beta,J)=\omega_{N,\beta,J}(H_N(\sigma,J))=
-\partial_{\beta}\log Z_N(\beta,J)=\partial_{\beta}\bigl(\beta
F_N(\beta,J)\bigr). 
\eqno(2.7)$$
The entropy $S_N(\beta,J)$ is
$$S_N(\beta,J)=-\sum_{\sigma_1\dots\sigma_N}(\rho_N \log
\rho_N)(\sigma;\beta,J),
\eqno(2.8)$$
with the usual bounds
$$0\le S_N(\beta,J)\le N \log2.
\eqno(2.9)$$These  variables are connected through the
second principle of thermodynamics
$$F_N(\beta,J)=U_N(\beta,J)-\beta^{-1}S_N(\beta,J).
\eqno(2.10)$$
All thermodynamic variables are random with respect to the external
noise $J$. However, we can also introduce the free energy per site
$f_N(\beta,J)$, and correspondently the internal energy per site
$u_N(\beta,J)$, and the entropy per site $s_N(\beta,J)$, such that
$$\eqalignno{ f_N(\beta,J)=N^{-1}F_N(\beta,J),\quad
u_N(\beta,J)&=N^{-1}U_N(\beta,J),\quad
s_N(\beta,J)=N^{-1}S_N(\beta,J),\cr
0\le s_N(\beta,J)\le  \log2,\quad
f_N(\beta,J)&=u_N(\beta,J)-\beta^{-1}s_N(\beta,J).
&(2.11)\cr}$$
On physical grounds, it is expected that these variables have a
good thermodynamic behavior in the limit when $N\to\infty$, in the
sense that they should converge almost surely, with respect to the
external $J$ noise, {\sl i.e.} with the possible exclusion of a set
of $J$ samples of zero measure in the $J$ probability space,
$$f_N(\beta,J)\to f_N(\beta),\quad u_N(\beta,J)\to
u_N(\beta),\quad s_N(\beta,J)\to s_N(\beta).
\eqno(2.12)$$
Alternatively, one can consider also convergence of the quenched
averages with respect the external noise
$$E\bigl(f_N(\beta,J)\bigr)\to f_N(\beta),\quad E\bigl(u_N(\beta,J)\bigr)\to
u_N(\beta),\quad E\bigl(s_N(\beta,J)\bigr)\to s_N(\beta).
\eqno(2.13)$$

For the sake of convenience, let us define the quenched
average
$$\alpha_N(\beta)=N^{-1}E\bigl(\log
Z_N(\beta,J)\bigr)=-\beta
E\bigl(f_N(\beta,J)\bigr). 
\eqno(2.14)$$
Clearly, $\alpha_N(\beta)$ is convex and increasing in $\beta$,
for $\beta\ge0$. Then we have the following elementary estimates
(see for example [6]).
\smallskip\noindent
{\bf Proposition 1.} For the previously defined
$\alpha_N(\beta)$ the following bounds hold, uniformly in $N$,
$$\eqalignno{ \log2\le\alpha_N(\beta)\le&
\log2+{\beta^2/4},\quad\hbox{for}\
\beta\le\bar{\beta}=2\sqrt{\log2}=1.665\dots,\cr
\le&\beta\sqrt{\log2},\quad\quad\hbox{for}\
\beta\ge\bar{\beta}.
&(2.15)\cr}$$

The proof (see for example [6]) is based on the following
simple, but important, inequality, relating quenched averages
with annealed averages
$$E\bigl(\log Z_N(\beta,J)\bigr)\le \log E\bigl( Z_N(\beta,J)\bigr).
\eqno(2.16)$$
In the annealed average the $J$'s participate to the thermodynamic
equilibrium. A simple calculation shows 
$$\eqalignno{E(Z_N(\beta,J))=&\sum_{\sigma_1\dots\sigma_N}\prod_{(i,j)}
E\bigl(\exp({\beta\over\sqrt{N-1}}J_{ij}\sigma_i\sigma_j)\bigr)\cr
&=
2^{N}(\exp{\beta^{2}\over{2(N-1)}})^{N(N-1)\over2}=(2\exp{\beta^{2}\over4})^N.
&(2.17)\cr}$$
Therefore, for any $\beta$
$$\alpha_N(\beta)\le \log2+{\beta^2/4}.
\eqno(2.18)$$
On the other hand, the lower bound in (2.15) is trivial, since convexity gives
$$Z_N(\beta,J)=\sum_{\sigma_1\dots\sigma_N}\exp(-\beta
H_N(\sigma,J))\ge 2^{N}\exp(-{\beta\over2^{N}}\sum_{\sigma_1\dots\sigma_N}
H_N(\sigma,J))=2^{N}.
\eqno(2.19)$$
As shown in [6], a simple argument, based on thermodynamic stability and the
bound (2.18), gives the upper bound in (2.15) for $\beta \ge \bar\beta$. In
fact, introduce the function
$$\phi(\beta)=\log2+\beta^2/4,
\eqno(2.20)$$
and define $\bar\beta$ such that, with
$\phi'(\beta)=(\partial_{\beta}\phi)(\beta)$,
$$\phi(\bar\beta)-\bar{\beta}\phi'(\bar\beta).
\eqno(2.21)$$
Therefore
$$\bar{\beta}=2\sqrt{\log2},\quad \phi(\bar\beta)=2 \log2,\quad
\phi'(\bar\beta)=\sqrt{\log2}.
\eqno(2.22)$$
Assume by absurd that there exists a point $\beta>\bar\beta$ where the upper
bound is violated, {\sl i.e.}
$$\alpha_N(\beta)> \beta\sqrt{\log2},\quad \beta>\bar\beta.
\eqno(2.23)$$
Notice that by convexity of $\alpha_N$ we would have at this point
$$\alpha_{N}'(\beta)\ge
(\beta-\bar\beta)^{-1}\bigl(\alpha_N(\beta)-\alpha_N(\bar\beta)\bigr)>
(\beta-\bar\beta)^{-1}(\beta\sqrt{\log2}-2\log2)=\sqrt{\log2}.
\eqno(2.24)$$
On the other hand, the average entropy is given by
$$E\big(s_N(\beta,J)\bigr)=\alpha_N(\beta)-\beta\alpha_{N}'(\beta),
\eqno(2.25)$$
as a consequence of (2.7,10,11,14). Therefore, by exploiting again convexity
and the bounds (2.23) in $\beta$ and (2.18) in $\bar\beta$, we would get
$$\eqalignno{
E\big(s_N(\beta,J)\bigr)&\le\alpha_N(\beta)-
\beta(\beta-\bar\beta)^{-1}\bigl(\alpha_N(\beta)-\alpha_N(\bar\beta)\bigr)\cr  
&=-\bar\beta(\beta-\bar\beta)^{-1}\alpha_N(\beta)+
\beta(\beta-\bar\beta)^{-1}\alpha_N(\bar\beta)\cr
&<-\bar\beta(\beta-\bar\beta)^{-1}\beta{\sqrt{\log2}}+
\beta(\beta-\bar\beta)^{-1}2\log2=0,
&(2.26)\cr}$$ 
which is impossible.

Of course, the bound given by Proposition 1 extends to the limit $N\to\infty$.
\smallskip\noindent
{\bf Proposition 2.} In the thermodynamic limit we have
$$\eqalignno{ \limsup_{N\to\infty}\alpha_N(\beta)\le&
\log2+{\beta^2/4},\quad\hbox{for}\
\beta\le\bar{\beta}=2\sqrt{\log2}=1.665\dots,\cr
\le&\beta\sqrt{\log2},\quad\quad\hbox{for}\
\beta\ge\bar{\beta}.
&(2.27)\cr}$$
As remarked in [6], these bounds show a phase transition for some
$\beta_c\le\bar\beta=2\sqrt{\log2}$, in the sense that annealing of the noise
$J$ gives wrong results in the limit  $N\to\infty$. On the other hand, next Theorem
3 shows that annealing is correct in the thermodynamic limit for any
$\beta\le1$. Therefore, the transition must occur at some critical point
$\beta_c$, such that $1\le\beta_c\le2\sqrt{\log2}$. On physical grounds (see
for example [2]), it is expected that $\beta_c=1$. This is also confirmed by
the buildig up of strong fluctuations in the $N\to\infty$ limit of global
thermodynamic variables, as $\beta\uparrow1$, shown in the beautyful and
detailed analysis of Aizenman, Lebowitz and Ruelle [6], based on graph
expansions.

\beginsection{3. THERMODYNAMIC LIMIT IN THE HIGH TEMPERATURE REGIME.}

Let us introduce the important order parameter $M^2_N(\beta)$ (see for
example [2],[6],[8])  
$$M^2_N(\beta)=
{2\over{N(N-1)}}\sum_{(i,j)}E\bigl(\omega^2_{N,\beta,J}(\sigma_i\sigma_j)\bigr),
\quad 0\le M^2_N(\beta)\le1,
\eqno(3.1)$$
and write the average internal energy as 
$$-E\bigl(u_N(\beta,J)\bigr)=\alpha'_N(\beta)=
{\beta\over2}\bigl(1-M^2_N(\beta)\bigr).
\eqno(3.2)$$
This comes from a simple integration by parts on the external Gaussian noise,
expressed in the form, reminiscent of the Wick theorem in quantum field theory,
$$E\bigl(J_{ij}F(J)\bigr)=E\bigl({\partial\over{\partial J_{ij}}}F(J)\bigr),
\eqno(3.3)$$
for any smooth function $F$ of the noise.

In fact, we can write, with obvious shorthand notations,
$$\eqalignno{-E\bigl(u_N(\beta,J)\bigr)&=
-N^{-1}E\bigl(\omega_{N}(H_N)\bigr)\cr
&=(N\sqrt{N-1})^{-1}\sum_{(i,j)}
E\bigl(J_{ij}\omega_{N}(\sigma_i\sigma_j)\bigr)\cr
&=\beta N^{-1}(N-1)^{-1}
\sum_{(i,j)}E\bigl(\omega_{N}(\sigma_i\sigma_j,\sigma_i\sigma_j)\bigr)\cr
&=\beta N^{-1}(N-1)^{-1}
\sum_{(i,j)}\Bigl(1-E\bigl(\omega_{N}^2(\sigma_i\sigma_j)\bigr)\Bigr).
&(3.4)\cr}$$
We have exploited the general expression 
$${\partial\over{\partial J_{ij}}}\omega_{N}(A)=
(\beta/\sqrt{N-1})\omega_{N}(A,\sigma_i\sigma_j),
\eqno(3.5)$$ 
where $\omega_{N}$ is the Boltzmann state $\omega_{N,\beta,J}$, and
$\omega(A,B)$ denotes truncation, {\sl i.e.}
$\omega(A,B)=\omega(AB)-\omega(A)\omega(B)$.

Then we can state the following\smallskip\noindent
{\bf Theorem 3.} With the definitions (2.14) and (3.1) we have
$$\eqalignno{\lim_{N\to\infty}\alpha_N(\beta)&=\log2+\beta^2/4,&
(3.6)\cr
 \lim_{N\to\infty}M^2_N(\beta)&=0, 
&(3.7)\cr}$$
uniformly for $0\le\beta\le1$.

Here we give a streamlined proof, which exploits only elementary thermodynamic
properties and positivity of \fl. We find convenient to split the proof in a
long series of simple statements. It can be also understood as an elementary
application of the cavity method.

Let us define 
$$\alpha_N(\beta,t)=N^{-1}E\bigl(\log\sum_{\sigma_1\dots\sigma_N}
\exp({\beta\over\sqrt{N-1}}\sum_{(i,j)}J_{ij}\sigma_i\sigma_j)
\exp({t\over\sqrt{N}}\sum_{i}J_{i}\sigma_i)\bigr), 
\eqno(3.8)$$
where we have introduced an additional parameter $t$, $t\ge0$, and an
additional fresh noise $J_i$, $i=1,2,\dots,N$, with the same properties of the
noise $J_{ij}$. Notice that $\alpha_N(\beta,0)=\alpha_N(\beta)$, as defined in
(2.14). Clearly, $\alpha_N(\beta,t)$ is convex and increasing in $\beta$ and
$t$, separately. For the sake of simplicity, we call $\omega_t$ the state
associated to the new {\sl Boltzmannfaktor}. When $t=0$, $\omega_t$ reduces to
the even Boltzmann state $\omega_N$.

Let us introduce the order parameters   
$$\eqalignno{M^2_N(\beta,t)&=
{2\over{N(N-1)}}\sum_{(i,j)}E\bigl(\omega^2_t(\sigma_i\sigma_j)\bigr),
\quad 0\le M^2_N(\beta,t)\le1,\cr
{\overline M}^2_N(\beta,t)&=
N^{-1}\sum_i E\bigl(\omega^2_t(\sigma_i)\bigr),
\quad 0\le {\overline M}^2_N(\beta,t)\le1,
&(3.9)\cr}$$
such that
$$\eqalignno{\partial_{\beta}\alpha_N(\beta,t)&=
{\beta\over2}\bigl(1-M^2_N(\beta,t)\bigr),\cr
\partial_{t}\alpha_N(\beta,t)&=
{t\over N}\bigl(1-{\overline M}^2_N(\beta,t)\bigr). 
&(3.10)\cr}$$ 

Notice that $M^2_N(\beta,0)=M^2_N(\beta)$ as in (3.1), while
${\overline M}^2_N(\beta)=0$, since $\omega_N$ is even. In a sense, the parameter
$t$ gives a smooth interpolation between a system with $N$ sites, at $t=0$,
and a system with $N+1$ sites, at $t=\sqrt{N/(N-1)}$, with a small change in
the temperature. The parameter $t$ denotes the strength of the coupling
between the original $N$ spins and an additional $(N+1)$th spin added, with
$J_i$ interpreted as $J_{i,N+1}$, $i=1,2,\dots,N$. A precise statement is
given by\smallskip\noindent
{\bf Lemma 4.} The following equalities hold    
$$\eqalignno{M^2_N(\beta\sqrt{(N-1)/N},\beta)&={\overline
M}^2_N(\beta\sqrt{(N-1)/N},\beta)=M^2_{N+1}(\beta),\cr
(N+1)\alpha_{N+1}(\beta)&=\log2+N\alpha_N(\beta\sqrt{(N-1)/N},\beta). 
&(3.11)\cr}$$
For the proof let us notice that we have 
$M^2_N(\beta,t)=E\bigl(\omega^2_t(\sigma_1\sigma_2)\bigr)$,
${\overline M}^2_N(\beta,t)=E\bigl(\omega^2_t(\sigma_1)\bigr)$,
$M^2_{N+1}(\beta)=E\bigl(\omega^2_{N+1}(\sigma_1\sigma_{N+1})\bigr)=
E\bigl(\omega^2_{N+1}(\sigma_1\sigma_2)\bigr)$, due to the complete symmetry
among the $J$'s. In fact, for each of the alternatives in
$(\dots,\dots,\dots)$, we can write 
$$\sum_{\sigma_1\dots\sigma_N}(1,\sigma_1,\sigma_1\sigma_2)
\exp({\beta\over\sqrt N}\sum_{(i,j)}^{1\dots
N}J_{ij}\sigma_i\sigma_j) \exp({\beta\over\sqrt{N}}\sum_{i}^{N}J_{i}\sigma_i)=$$
$$=
{1\over2}\sum_{\sigma_1\dots\sigma_N \sigma_{N+1}}\dots=
{1\over2}\sum_{\sigma_1\dots\sigma_{N+1}}
(1,\sigma_1\sigma_{N+1},\sigma_1\sigma_2)\exp({\beta\over\sqrt{N}}
\sum_{(i,j)}^{1\dots N+1}J_{ij}\sigma_i\sigma_j),
\eqno(3.12)$$
where we have added the dummy summation variable $\sigma_{N+1}$ in the first
equality, and performed the change of variables
$\sigma_i\to\sigma_i\sigma_{N+1}$, $i=1,2,\dots,N$, in the second equality.

In the following  theorem we exploit the mild dependence of
$\alpha_N(\beta,t)$ on $t$, as a consequence of the $1/N$ factor in the $t$
derivative in (3.10), and simple convexity estimates, in order to have bounds
on $M^2_N(\beta,t)$, uniform in $t$.\smallskip\noindent
{\bf Theorem 5.} Assume $\beta>\Delta\beta>0$, then we have, uniformly for
$0\le t\le\tilde\beta$,  
$$M^2_N(\beta+\Delta\beta)-{\Delta\beta\over\beta}-
{{\tilde\beta}^2\over{N\beta\Delta\beta}} \le M^2_N(\beta,t) \le
M^2_N(\beta-\Delta\beta)+{\Delta\beta\over\beta}+
{{\tilde\beta}^2\over{N\beta\Delta\beta}}. 
\eqno(3.13)$$
For the proof, let us notice that (3.10) gives
$$0\le\partial_{t}\alpha_N(\beta,t)\le
{t\over N},\quad 0\le\alpha_N(\beta,t)-\alpha_N(\beta)\le 
{t^2\over {2N}}.
\eqno(3.14)$$
Therefore, by convexity we can write
$$\eqalignno{\Delta\beta\
\alpha'_N(\beta,t)&\ge\alpha_N(\beta,t)-\alpha_N(\beta-\Delta\beta,t)\cr
&\ge
\alpha_N(\beta)-\alpha_N(\beta-\Delta\beta)-{t^2\over
{2N}}\cr
&\ge\Delta\beta\ \alpha'_N(\beta-\Delta\beta)-
{{\tilde\beta}^2\over{2N}}.
&(3.15)\cr}$$
>From this and (3.10) we have
$$\alpha'_N(\beta,t)\ge\alpha'_N(\beta-\Delta\beta)-
{{\tilde\beta}^2\over{2N\Delta\beta}},$$
$${\beta\over2}\bigl(1-M^2_N(\beta,t)\bigr)\ge
{{\beta-\Delta\beta}\over2}\bigl(1-M^2_N(\beta-\Delta\beta)\bigr)-
{{\tilde\beta}^2\over{2N\Delta\beta}}.
\eqno(3.16)$$
We conclude
$$M^2_N(\beta,t)\le(1-{\Delta\beta\over\beta})M^2_N(\beta-\Delta\beta)+
{\Delta\beta\over\beta}+{{\tilde\beta}^2\over{N\beta\Delta\beta}}.
\eqno(3.17)$$
This gives the upper bound in the theorem, and the lower bound is
proven analogously.

Through a straigthforward calculation and integration by parts, we can now
prove the following basic theorem, which gives the $t$ derivative of the order
parameter ${\overline M}^2_N(\beta,t)$ in terms of the order parameter
$M^2_N(\beta,t)$ and some \fl\ of variables of the theory (see also
[8]).\smallskip\noindent
{\bf Theorem 6.} The following equality holds 
$$\eqalignno{\partial_{t^2}\Bigl(N^{-1}\sum_i
E\bigl(\omega^2_t(\sigma_i)\bigr)\Bigr)&=
N^{-2}\sum_{i,j}E\bigl(\omega^2_t(\sigma_i\sigma_j)\bigr)\cr-
E\Bigl(\big(N^{-1}\sum_i\omega^2_t(\sigma_i)\bigr)^2\Bigr)&-
4E\big(\omega_t(A,A)\bigr), &(3.18)\cr}$$
where $A$ is the variable
$$A=N^{-1}\sum_i\sigma_i\omega_t(\sigma_i).
\eqno(3.19)$$
For the proof, let us evaluate
$$\partial_{t}\Bigl(N^{-1}\sum_i E\bigl(\omega^2_t(\sigma_i)\bigr)\Bigr)=
2 N^{-1}\sum_i E\bigl(\omega_t(\sigma_i)\partial_{t}\omega_t(\sigma_i)\bigr)=$$
$$=
2 (N\sqrt N)^{-1}\sum_{i,j}
E\bigl(J_j\omega_t(\sigma_i)\omega_t(\sigma_i,\sigma_j)\bigr)=$$
$$=
2 t N^{-2}\sum_{i,j}
E\bigl(\omega_t(\sigma_i,\sigma_j)+
\omega_t(\sigma_i)\omega_t(\sigma_i,\sigma_j,\sigma_j)\bigr),
\eqno(3.20)$$
where we have exploited
$$\partial_{t}\omega_t(\sigma_i)=
{1\over\sqrt N}\sum_j J_j\omega_t(\sigma_i,\sigma_j),\quad
{\partial\over{\partial J_j}}\omega_t(\sigma_i)=
{t\over\sqrt N}\omega_t(\sigma_i,\sigma_j),$$
$$
{\partial\over{\partial J_j}}\omega_t(\sigma_i,\sigma_j)=
{t\over\sqrt N}\omega_t(\sigma_i,\sigma_j,\sigma_j).
\eqno(3.21)$$
Now we consider the identities
$$\eqalignno{\omega_t(\sigma_i,\sigma_j)&=\omega_t(\sigma_i\sigma_j)-
\omega_t(\sigma_i)\omega_t(\sigma_j),\cr
\omega_t(\sigma_i,\sigma_j,\sigma_j)&=
-2 \omega_t(\sigma_i,\sigma_j) \omega_t(\sigma_j).
&(3.22)\cr}$$
By collecting all terms, we can write (3.20) in the form
$$\partial_{t^2}\Bigl(N^{-1}\sum_i E\bigl(\omega^2_t(\sigma_i)\bigr)\Bigr)=
N^{-2}\sum_{i,j}
E\bigl(\omega_t^2(\sigma_i\sigma_j)
\omega_t^2(\sigma_i)\omega_t^2(\sigma_j)-
4 \omega_t(\sigma_i)\omega_t(\sigma_j)\omega_t(\sigma_i,\sigma_j)\bigr),
\eqno(3.23)$$
and the theorem is proven.

Now we exploit the obvious positivity inequalities
$$\omega_t(A,A)\ge0,\quad
E\Bigl(\big(N^{-1}\sum_i\omega^2_t(\sigma_i)\bigr)^2\Bigr)\ge
E\Bigl(\big(N^{-1}\sum_i\omega^2_t(\sigma_i)\bigr)\Bigr)^2,
\eqno(3.24)$$
and the inequality
$$\eqalignno{N^{-2}\sum_{i,j}E\bigl(\omega^2_t(\sigma_i\sigma_j)\bigr)&=
N(N-1)N^{-2}E\bigl(\omega^2_t(\sigma_1\sigma_2)\bigr)+{1\over N}\cr
&=
(1-{1\over N})M^2_N(\beta,t)+{1\over N}\le
M^2_N(\beta,t)+{1\over N},
&(3.25)\cr}$$
together with the definitions (3.9), in order to derive\smallskip\noindent
{\bf Theorem 7.} The following inequalities hold
$$\eqalignno{\partial_{t^2}{\overline M}^2_N(\beta,t)+{\overline M}^4_N(\beta,t)&\le
M^2_N(\beta,t)+{1\over N},\quad \hbox{for any $t$},\cr
&\le {\widetilde M}^2_N(\beta,\tilde\beta),\quad\hbox{uniformly for $0\le
t\le\tilde\beta$},   
&(3.26)\cr}$$
where
$${\widetilde M}^2_N(\beta,\tilde\beta)=
\max_{0\le t\le\tilde\beta}M^2_N(\beta,t)+{1\over N},
\eqno(3.27)$$
$${\overline M}^2_N(\beta,t)\le
{\widetilde M}_N(\beta,\tilde\beta)\tanh\big(t^2{\widetilde
M}_N(\beta,\tilde\beta)\bigr),\quad\hbox{uniformly for $0\le
t\le\tilde\beta$}.  
\eqno(3.28)$$

In fact, (3.26) follows from (3.18), (3.24), (3.25) and the definition (3.27),
while (3.28) follows from a simple integration.

Let us now put $\beta\sqrt{(N-1)/N}$ in place of $\beta$ in (3.28), and then 
$\tilde\beta=\beta$, and $t=\beta$. We get  
$$\eqalignno{M^2_N&(\beta\sqrt{(N-1)/N},\beta)={\overline
M}^2_N(\beta\sqrt{(N-1)/N},\beta)\cr&\le
{\widetilde M}_N(\beta\sqrt{(N-1)/N},\beta)\tanh\big(t^2{\widetilde
M}_N(\beta\sqrt{(N-1)/N},\beta)\bigr).
&(3.29)\cr}$$
Now we are ready for the proof of the following key theorem.\smallskip\noindent
{\bf Theorem 8.} Define
$$M^2_N=\max_{0\le\beta\le1}M^2_N(\beta),\quad M^2_N\le1, 
\eqno(3.30)$$
then for large $N$
$$M^4_N\le3{\biggl({{e^2+1}\over{e^2-1}}\biggr)}^2 ({4\over\sqrt{N-1}}+{1\over N}),
\eqno(3.31)$$
hence
$$\lim_{N\to\infty}M^2_N=0.
\eqno(3.32)$$

For the proof, let us notice that the definitions (3.27) and (3.30), and the
uniform upper bound (3.13) of Theorem 5 give for $\beta\le1$
$$\eqalignno{{\widetilde M}^2_N(\beta\sqrt{(N-1)/N},\beta)&\le
M^2_N(\beta\sqrt{(N-1)/N}-\Delta\beta)+
{\Delta\beta\over\beta}{\sqrt N\over\sqrt{N-1}}+
{\beta\over\Delta\beta}{1\over N}{\sqrt N\over\sqrt{N-1}}+{1\over N}\cr
&\le
M^2_N+{1\over\sqrt{N-1}}(
{{\Delta\beta\sqrt N}\over\beta}+
{\beta\over{\Delta\beta\sqrt N}})+{1\over N}.
&(3.33)\cr}$$
It is convenient to put $\Delta\beta=\beta/\sqrt N$, so that
$${\widetilde M}^2_N(\beta\sqrt{(N-1)/N},\beta)\le
M^2_N+{2\over\sqrt{N-1}}+{1\over N}.
\eqno(3.34)$$
But in the same conditions we derive from the lower bound (3.13)
$${\widetilde M}^2_N(\beta\sqrt{(N-1)/N},\beta)\ge
M^2_N\big(\beta(\sqrt{(N-1)/N}+{1\over\sqrt{N}})\big)-{2\over\sqrt{N-1}}.
\eqno(3.35)$$
Therefore, from (3.29), (3.34) and (3.35) we derive
$$\eqalignno{M^2_N\big({\beta\over\sqrt{N}}(1+\sqrt{N-1})\big)&\le
{2\over\sqrt{N-1}}+{M'}_N \tanh (\beta^2 {M'}_N),\cr 
{M'}_N&=\sqrt{M^2_N+{2\over\sqrt{N-1}}+{1\over N}}.
&(3.36)\cr}$$
Notice that $(1+\sqrt{N-1})/\sqrt{N}>1$, therefore, if we take the maximum in
(3.36), for $0\le\beta\le1$, we immediately have the following important bound
$${M'}^2_N\le
{4\over\sqrt{N-1}}+{1\over N}+{M'}_N \tanh (\beta^2 {M'}_N),
\eqno(3.37)$$
where we added ${2/\sqrt{N-1}}+{1/N}$ to both members for
convenience. Since the function $x(x-\tanh x)$ is strictly
increasing for $x\ge0$, and $M_N\le {M'}_N$, we have also, 
$$M_N(M_N-\tanh M_N)\le{4\over\sqrt{N-1}}+{1\over N},
\eqno(3.38)$$
which shows that $M_N\to0$ as $N\to\infty$.

In order to have an estimate on the rate of convergence, we exploit the
following inequality, holding for $0\le x\le1$, 
$$x(x-\tanh x)\ge x^4(\tanh1)^2/3,
\eqno(3.39)$$
and the theorem follows.

Finally, we can prove Theorem 3. In fact, from $M_N^2(\beta)\le M_N$, for 
$0\le\beta\le1$, we have the uniform convergence in (3.7). On the other hand,
we also have  
$$\eqalignno{\alpha'_N(\beta)&=
{\beta\over2}\bigl(1-M^2_N(\beta)\bigr)\ge
{\beta\over2}(1-M^2_N),\cr
\alpha_N(\beta)&\ge
\log2+\beta^2(1-M^2_N)/4,
&(3.40)\cr}$$
and the uniform convergence in (3.6) follows.

In conclusion, we see that the validity of the annealed approximation holds
uniformly in  $0\le\beta\le1$, in the infinite
volume limit, as a simple consequence of
thermodynamic stability and positivity of the
mean square \fl.

\beginsection{MARGINAL MARTINGALE AND CAVITY METHODS.}

Let $\omega$ be a generic even state on the Ising variables $\sigma_1,\dots,\sigma_N$, possibly
depending on a stale noise $J_{ij}$. Introduce the marginal martingale function $\psi_N(\omega,t)$
defined by 
$$\psi_N(\omega,t)=E\log\omega(\exp{t\over\sqrt N}\sum_i J_i\sigma_i),
\eqno(4.1)$$
where $J_i$ are a fresh noise, as in (3.8). Since $\omega$ is even, we can
substitute the $\cosh$ in place of the $\exp$ in (4.1).

We have the bounds given by
\smallskip
\noindent
{\bf Theorem 9.} For any   $t$, the following bounds hold uniformly in $\omega$
$$\int\log\cosh(tz)\ d\mu(z)\le\psi_N(\omega,t)\le N\int\log\cosh(tz/\sqrt N)\ d\mu(z),
\eqno(4.2)$$
where $d\mu(z)=\exp(-z^2/2)\ dz/\sqrt{2\pi}$ is the unit Gaussian distribution. The upper bound is
realized if $\omega$ in (4.1) is taken as the symmetric product state
$\omega^{(0)}$, where all configurations of the $\sigma$'s have the same
probability $2^{-N}$. This is the case if $\omega$ is the Boltzmann state
$\omega_N$ at $\beta=0$. The lower bound is realized if $\omega$ in (4.1) is a
state $\overline{\omega}$ which gives equal weigths $1\over2$ to some fixed
configurations  $\bar{\sigma_1},\dots,\bar{\sigma_N}$ and its inverted
one  $-{\bar\sigma}_1,\dots,-{\bar\sigma}_N$, and zero to all other
configurations.

Proof. First of all let us show that the bounds are in fact equalities for some states. For the
upper bound let $\omega$ in (4.1) be the symmetric product state $\omega^{(0)}$.
In this case we have     
$$\log\omega^{(0)}(\exp{t\over\sqrt N}\sum_i J_i\sigma_i)=
\sum_i\log\omega^{(0)}(\exp{t\over\sqrt N} J_i\sigma_i)=
\sum_i\log \cosh ({t\over\sqrt N} J_i). 
\eqno(4.3)$$
By taking the average $E$ and exploiting the symmetry in the $J$'s, we have the upper bound in (4.2)
as an equality. For the lower bound, let $\omega$ be the state $\overline{\omega}$ which gives
probability $1\over2$ to each of the two configurations $\sigma_i=1$, and $\sigma_i=-1$  for all
$i$'s, for example. In this case we have
$$\overline{\omega}(\exp{t\over\sqrt N} J_i\sigma_i)=
\cosh ({t\over\sqrt N} J_i).
\eqno(4.4)$$
By taking the $E$ average we get the lower bound in (4.2) as an equality.

Let us now prove the bounds for a generic state. Introduce the additional Ising variables
$\epsilon_1,\dots,\epsilon_N$, with symmetric distribution, and let $E'$ denote the related average.
Since the $J$'s have symmetric distributions, by annealing in $E'$, we have 
$$\eqalignno{E\log\omega(\exp{t\over\sqrt N}\sum_i J_i\sigma_i)&=
EE'\log\omega(\exp{t\over\sqrt N}\sum_i J_i\sigma_i\epsilon_i)\cr
&\le E\log\omega E'(\exp{t\over\sqrt N}\sum_i J_i\sigma_i\epsilon_i)\cr
&=\sum_i\log \cosh ({t\over\sqrt N} J_i), 
&(4.5)\cr}$$
where we have freely exchanged $E'$ and $\omega$. Therefore, we are reduced to (4.3), and the upper
bound follows. Obviously, we could anneal completely the $J$ variables and get the slightly weaker
bound uniform in $N$
$$E\log\omega(\exp{t\over\sqrt N}\sum_i J_i\sigma_i)\le{1\over2}t^2.
\eqno(4.6)$$
For the lower bound, it is enough to write
$$E\log\omega(\cosh{t\over\sqrt N}\sum_i J_i\sigma_i)\ge
\omega(E\log\cosh{t\over\sqrt N}\sum_i J_i\sigma_i),
\eqno(4.7)$$
by quenching also the variables $\sigma$, in some sense. In the $E$ average the variables
$\sigma$ do not play any role, and the lower bound follows.

Let us remark that the upper bound in (4.2) is realized for the state
$\omega^{(0)}$ of maximum entropy $N\log2$, while the lower bound corresponds to
states $\overline{\omega}$ of minimum entropy $\log2$, compatible with eveness.
It would be interesting to find the appropriate bounds if $\omega$ has some
given fixed entropy.

The interest in the introduction of the marginal martingale relies on the possibility to derive
information on the thermodynamic variables from information on $\psi_N(\omega,t)$, for some
properly chosen state. Let us remark that $\psi_N$ and related quantities play also a central
role in the deep study of \fl\ made by Pastur and Scherbina [8,9]. We refer to [10] for an extensive
treatment of the marginal martingale. Here we give only some typical results.

Let us define the particular marginal martingale $\psi_N(\beta)$, associated to the mean field
spin glass model, by replacing the generic state $\omega$ in (4.1) with
the Boltzmann state $\omega'_N$ at temperature $\beta\sqrt{(N-1)/N}$,
and putting $t=\beta$,       
$$\psi_N(\beta)=\psi_N(\omega'_N,t)=
N\bigl(\alpha_{N+1}(\beta\sqrt{{N-1}\over N},\beta)-
\alpha_N(\beta\sqrt{{N-1}\over N})\bigr).
\eqno(4.8)$$ 
>From (3.11) we have
$$(N+1)\alpha_{N+1}(\beta)=
\log2+N\alpha_N(\beta\sqrt{{N-1}\over N})+\psi_N(\beta).
\eqno(4.9)$$
We see that $\psi_N(\beta)$ connects the free energy for a system with $N+1$
particles with the free energy  for a system with $N$ particles, with a small
change in the temperature. This is a typical feature of the cavity method (see
for example [4]).

The following theorem shows how information on the limiting behavior of  
$\psi_N(\beta)$ translates into information on the behavior of the
thermodynamic variables.
\smallskip\noindent
{\bf Theorem 10.} Assume that a lower bound of the type
$$\psi_N(\beta)\ge{\overline\psi}_N(\beta)
\eqno(4.10)$$
holds uniformly for $0\le\beta\le\tilde\beta$, $N\ge K$. Then we have
$$\liminf_{N\to\infty}\alpha_N(\beta)\ge
\log2+\int^1_0{\overline\psi}(\beta\sqrt{1-q})\ dq,
\eqno(4.11)$$
for $0\le\beta\le\tilde\beta$. An analogous statement holds for an upper bound.
\hfill\break
Assume that the following limit exists
$$\lim_{N\to\infty}\psi_N(\beta)=\psi(\beta),
\eqno(4.12)$$
uniformly on a compact region $0\le\beta\le\tilde\beta$, with $\psi(\beta)$
continuous in $\beta$, as a consequence. Let us define  
$$\eqalignno{\alpha(\beta)&=
\log2+\int^1_0\psi(\beta\sqrt{1-q})\ dq\cr
&=\log2+\beta^{-2}\int^{\beta^2}_0\psi(\beta')\ d{\beta'}^2,
&(4.13)\cr}$$
so that the $\beta$ derivative $\alpha'(\beta)$ exists and the following holds
$$\alpha(\beta)+\beta\alpha'(\beta)/2=\log2+\psi(\beta).
\eqno(4.14)$$
Then we have, for $0\le\beta\le\tilde\beta$, 
$$\lim_{N\to\infty}\alpha_N(\beta)=\alpha(\beta),
\lim_{N\to\infty}\alpha'_N(\beta)=\alpha'(\beta),
\lim_{N\to\infty}\bigl((N+1)\alpha_{N+1}(\beta)-N\alpha_N(\beta)\bigr)=\alpha(\beta).
\eqno(4.15)$$
Moreover, $\psi(\beta)$ and $\psi(\beta)/\beta$ are increasing in $\beta$, and 
$$\psi(\beta)=\beta^2/2,
\eqno(4.16)$$
in the fluid region $0\le\beta\le1$.

We give only a sketch of the proof, by referring to [10] for complete details.
Let us iterate (4.9) $N-K$ times  
$$\eqalignno{(N+1)\alpha_{N+1}(\beta)&=
(N-K+1)\log2+K\alpha_K(\beta\sqrt{{K-1}\over N})\cr
&+\psi_N(\beta)+\psi_{N-1}(\beta\sqrt{{N-1}\over N})+\dots
+\psi_K(\beta\sqrt{K\over N}).
&(4.17)\cr}$$
Therefore, by assuming the uniform bound (4.10), we have
$$\alpha_{N+1}(\beta)\ge
{{N-K+1}\over{N+1}}+
{K\over{N+1}}\alpha_K(\beta\sqrt{{K-1}\over N})+
{1\over{N+1}}\sum^N_{K'=K}{\overline\psi}(\beta\sqrt{K'\over N}),
\eqno(4.18)$$
and derive (4.11) in the limit $N\to\infty$, where the discrete variable $K'/N$
becomes the continuous $1-q$. The other statements follow by standard arguments.

A simple application can be obtained by combining the lower bound in (4.2) with
(4.11).\smallskip\noindent
{\bf Theorem 11.} The following lower bound holds
$$\liminf_{N\to\infty}\alpha_N(\beta)\ge
\log2+\int^1_0 dq\int\log\cosh(\beta z\sqrt{1-q})\ d\mu (z).
\eqno(4.19)$$
This bound is a mild improvement over the analogous one found in [6] through a
clever choice of trial states in the variational principle for the free energy.
In fact, since $\log\cosh x$ is an even convex function of $x$, we have 
$$\int\log\cosh(\beta z\sqrt{1-q})\ d\mu (z)\ge
\log\cosh(\beta\sqrt{1-q}\int|z|\ d\mu (z))=
\log\cosh(\beta\sqrt{1-q}\sqrt{2/\pi}),
\eqno(4.20)$$
and the bound in [6] follows from (4.19).

\beginsection 5. THE FUNCTIONAL ORDER PARAMETER.

While the approximate solution of the model, given by Kirkpatrick and
Sherrington in [1], involves a numerical order parameter, function of the
temperature, it was soon realized by Parisi that the complete solution [3,5]
must be expressed through a functional order parameter $x$, which for a given
temperature depends on an auxiliary variable $q$, both $x$ and $q$ taking
values on the interval $[0,1]$. This was suggested by numerical computations
and a deep physical intuition about the rich structure of phases of the
system, in the infinite volume limit, for $\beta>1$.

Here we prove that any marginal martingale can be expressed through a
functional order parameter of Parisi type, also in the infinite volume limit
$N\to\infty$. Moreover, we show that there is a very large set, in the convex
space of functional order parameters, which gives equivalent results
in the expression of the marginal martingale. This kind of gauge freedom in
the choice of the functional order parameter raises the problem of picking
the most convenient representation, according to some criterion. Parisi
choice [2,3,5], fully discussed in [10] in the general frame introduced here,
is very natural, because it gives simple physical interpretations for the
parameters $x$ and $q$, and very useful for the applications. But, in
principle, other choices are also possible.

The main result of this Section is given by the following\smallskip\noindent
{\bf Theorem 12.} For a given even state $\omega$,  on the
Ising variables $\sigma_1,\dots,\sigma_N$, possibly depending on some stale noise,
let us introduce, as in (4.1), the marginal martingale 
$$\psi(\beta)=E\log\omega(\cosh{\beta\over\sqrt
N}\sum_i J_i\sigma_i). 
\eqno(5.1)$$
Then, for each $\omega$ and $\beta$, there exists a functional order parameter
$$x:\quad[0,1]\ni q\to x(q)\in[0,1],
\eqno(5.2)$$
such that
$$\psi(\beta)=f(0,0),
\eqno(5.3)$$
where $f(q,y)$, $0\le q\le1$, $y\in R$, is the solution of the nonlinear
antiparabolic equation 
$$(\partial_q f)(q,y)+{1\over2}\bigl(f^{\prime\prime}(q,y)+x(q){f^\prime}^2(q,y)\bigr)=0,
\eqno(5.4)$$
with final condition
$$f(1,y)=\log\cosh(\beta y).
\eqno(5.5)$$
In (5.4), $f^\prime=\partial_y f$ and $f^{\prime\prime}=\partial_y^2 f$.
A possible expression of $x$ is implicitely given by
$$\bigl(1-x(q)\bigr)E\bigl(\widetilde\omega({\tilde f}^{\prime2})\bigr)=
N^{-1}\sum_i E\bigl({\widetilde\omega}^2(\sigma_i{\tilde f}^\prime)\bigr),
\eqno(5.6)$$
where
$$\tilde f=f(q,\sqrt{q}{1\over\sqrt N}\sum_i J_i\sigma_i),\quad
\tilde f^\prime=f^\prime(q,\sqrt{q}{1\over\sqrt N}\sum_i J_i\sigma_i),
\eqno(5.7)$$
and $\widetilde\omega$ is the state 
$${\widetilde\omega}(A)=\omega(A\exp\tilde f)/\omega(\exp\tilde f).
\eqno(5.8)$$ 

The proof is very simple, and can be understood as a consistent correction to
successive annealing of the $J$ variables [10]. It also shows how naturally the
equation (5.4) arises from integration by parts. For a generic $f(q,y)$ let us
introduce the function $\phi$ defined by
$$\eqalignno{\phi:\quad[0,1]\ni q\to\phi(q)&=
E\log\omega\bigl(\exp f(q,\sqrt{q}{1\over\sqrt N}\sum_i J_i\sigma_i)\bigr)\cr
&=E\log\omega(\exp\tilde f).
&(5.9)\cr}$$
Notice the typical Brownian scaling $\sqrt q$ in the $y$ variable.

Assume for $f$ the boundary condition (5.5). Then we have
$$\phi(1)=\psi(\beta),\quad \phi(0)=f(0,0).
\eqno(5.10)$$
Let us now calculate the derivative of $\phi(q)$. We have 
$${d\over{dq}}\exp\tilde f=(\partial_q \tilde f)\exp\tilde f+
{1\over{2\sqrt q}}{1\over\sqrt N}\sum_i J_i\sigma_i\tilde f^\prime\exp\tilde f.
\eqno(5.11)$$
Therefore
$${d\over{dq}}\phi(q)=E\bigl({\widetilde\omega}(\partial_q \tilde f)\bigr)+
{1\over{2\sqrt q}}{1\over\sqrt N}\sum_i 
E\bigl(J_i{\widetilde\omega}(\sigma_i\tilde f^\prime)\bigr).
\eqno(5.12)$$
Now we integrate by parts on the fresh noise $J_i$, by using
$${\partial\over{\partial J_i}}\exp\tilde f=
\sqrt{q\over N}\sigma_i\tilde f^\prime\exp\tilde f,\quad
{\partial\over{\partial J_i}}\tilde f^\prime=
\sqrt{q\over N}\sigma_i\tilde f^{\prime\prime}.  
\eqno(5.13)$$
By collecting all terms, we have
$${d\over{dq}}\phi(q)=
E\bigl({\widetilde\omega}(\partial_q \tilde f)+{1\over2}\tilde f^{\prime\prime}+
{1\over2}{\tilde f}^{\prime2})\bigr)-
{1\over2}{1\over N}\sum_i 
E\bigl({\widetilde\omega}^2(\sigma_i{\tilde f^\prime})\bigr).
\eqno(5.14)$$
Therefore,  if $f$ is chosen so to satisfy (5.4) with $x$ given by (5.6), we
see immediately that $\phi(q)$ does not depend on q, and (5.3) follows from
(5.10). Finally, let us notice that the inequalities
$$0\le{\widetilde\omega}^2(\sigma_i{\tilde f^\prime})\le
{\widetilde\omega}({\tilde f}^{\prime2})
\eqno(5.15)$$
imply
$$0\le x(q)\le1,
\eqno(5.16)$$
and the theorem is fully proven.

We refer to [10] for a detailed discussion about the properties of the solution
$f(q,y;x)$
of the antiparabolic equation (5.4), with final condition (5.5), as a
functional of a generic given $x$, as in (5.2). Here we only state the following
\smallskip\noindent
{\bf Theorem 13.} The function $f$ is monotone in $x$, in the sense that 
$x(q)\le{\bar x}(q)$, for all $0\le q\le1$, implies 
$f(q,y;x)\le f(q,y;{\bar x})$, for any $0\le q\le1$, $y\in R$. Moreover $f$
is continuous in the $L^1(dq)$ norm. In fact, for generic $x$, $\bar x$, we
have    
$$|f(q,y;x)-f(q,y;{\bar x})|\le
{\beta^2\over2}\int^1_q|x(q')-{\bar x}(q')|\ dq'.
\eqno(5.17)$$

Let us now introduce the extremal order parameters $x_0(q)\equiv 0$ and
$x_1(q)\equiv1$, such that for any $x$ we have $x_0(q)\le x\le x_1(q)$. It is
simple to solve (5.4), (5.5) in these cases in the form
$$f_0(q,y)=\int\log\cosh \bigl(\beta(y+z\sqrt{1-q})\bigr)\ d\mu(z),\quad
f_1(q,y)=\log\cosh(\beta y)+\beta^2(1-q)/2.
\eqno(5.18)$$
Therefore, we have for the $f$ associated to a generic $x$ the bounds
$$f_0(q,y)\le f(q,y)\le f_1(q,y),
\eqno(5.19)$$
and in particular at the point $q=0$, $y=0$,
$$f_0(0,0)\equiv\int\log\cosh (\beta z)\ d\mu(z)\le
f(0,0)\le(\beta^2/2)\equiv f_1(0,0). 
\eqno(5.20)$$
We recognize that these bounds are the same as found in (4.2) and (4.6),
written for $t=\beta$. This remark leads to far reaching consequences.
\smallskip\noindent
{\bf Theorem 14} Let $\psi(\beta)$ be a marginal martingale for $N$
particles, or a limit for $N\to\infty$, therefore satisfying the bounds
$$\int\log\cosh (\beta z)\ d\mu(z)\le
\psi(\beta)\le\beta^2/2.
\eqno(5.21)$$
Let $x_{\epsilon}$ be a generic family of functional order parameters
depending continuously in the $L^1$ norm on the variable $\epsilon$,
$0\le\epsilon\le1$, with $x_0\equiv0$, and $x_1\equiv1$, and nondecreasing in
$\epsilon$. Then, there exists an $\epsilon(\beta)$ such that
$$\psi(\beta)=f(0,0;x_{\epsilon(\beta)}),
\eqno(5.22)$$
where $f$ is defined by (5.4), (5.5) with $x$ replaced by
$x_{\epsilon(\beta)}$.

The proof follows easily from the {\sl a priori} bounds and continuity. An
easy way to construct families of $x_{\epsilon}$ is the following. Let 
$\bar x$ be some fixed functional order parameter. Introduce
$x_{\epsilon}$, for $0\le\epsilon\le1$, defined by     
$$x_{\epsilon}(q)=[{\bar x}(q)-1+2\epsilon],
\eqno(5.23)$$
where $[\dots]$ denotes truncation outside of the interval $[0,1]$,
{\sl i.e.} $[y]$ takes the value $0$ if $y\le0$, $y$ if $0\le
y\le1$, and $1$ if $y\ge1$.

In conclusion, we have \smallskip\noindent
{\bf Theorem 15.} If $\psi(\beta)$ is a marginal martingale, then
for each $\beta$ there exists a nonempty hypersurface
$\Sigma_\beta$ of functional order parameters such that
$$\psi(\beta)=f(0,0;x),\quad\quad\hbox{for any}\quad
x\in\Sigma_\beta.
\eqno(5.24)$$

\beginsection 6. CONCLUSIONS AND OUTLOOK.

We have seen that many important properties of the mean field spin glass model
can be derived by elementary methods, based on the fluctuations with respect to
the external noise.

In particular, we have shown that the positivity of the mean square deviations
alone forces the annealed approximation, for the free energy, to be correct in
the thermodynamic limit, in the high temperature fluid region, up to the
critical point.

In general, the corrections to annealing can be easily expressed through a
functional order parameter, of the type introduced by Parisi in his ingenious
discoveries about the physical properties of the spin glass system.

The methods exploited in this paper can be easily extended to other mean field
disordered theories, with variables more general that the Ising variables, for
example continuous ones.

Finally, we refer to [10] for a more complete description of the martingale
method and its applications.\vfill\eject

\beginsection REFERENCES

\item{ [1]} D. Sherrington and S. Kirkpatrick: Solvable model of a spin glass,
Phys. Rev. Lett., {\bf35}, 1792 (1975).
\item{ [2]} M. M\'ezard, G. Parisi, and M. A. Virasoro: {\sl Spin Glass Theory
and Beyond}, World Scientific, Singapore, 1987, and reprints included there.
\item{ [3]} G. Parisi: A sequence of approximate solutions to the S-K
model for spin glasses, J. Phys. {\bf A13}, L-115 (1980).
\item{ [4]}  M. M\'ezard, G. Parisi, and M. A. Virasoro: SK Model: the
Replica Solution without Replicas, Europhys. Lett., {\bf1}, 77 (1986).
\item{ [5]} G. Parisi: The order parameter for spin glasses: A function on the
interval 0-1, J. Phys. {\bf A13}, L-1101 (1980).  
\item{ [6]} M. Aizenman, L. Lebowitz, and D. Ruelle: Some Rigorous
Results on the \sk\ Model of Spin Glasses, Commun. Math. Phys., {\bf 112},
3-20 (1987).
\item{ [7]} J. Fr\"ohlich and B. Zegarlinski: Some Comments on the \sk\
Model of Spin Glasses, Commun. Math. Phys., {\bf 112},
553 (1987).
\item{ [8]} L. A. Pastur and M. V. Shcherbina: The Absence of
Self-Averaging of the Order Parameter in the \sk\ Model, J. Stat. Phys., {\bf
62}, 1 (1991).
\item{ [9]} M. V. Scherbina: More about Absence of
Selfaverageness of the Order Parameter in the \sk\ Model, CARR Reports in
Mathematical Physics, n. 3/91, Department of Mathematics, University of Rome
``La Sapienza'', 1991.
\item{[10]} F. Guerra: On the mean field spin glass model, in preparation.
\vfill\eject\bye